\documentclass[12pt]{article}
\usepackage {pictex}

\def\putcircbar at #1 #2 with fuzz #3 {%
  \put {\Large $\circ$} at {#1} {#2} 
  \dimen0=\Ydistance{#3}
  \put{\vbox{\hsize=\crossbarlength
    \hrule height \linethickness
    \vskip -.5\linethickness
    \centerline{\vrule width \linethickness height 2\dimen0}
    \nointerlineskip
    \vskip -.5\linethickness
    \hrule height \linethickness}} at {#1} {#2} }
   
\newdimen\xposition
\newdimen\yposition
\newdimen\dyposition
\newdimen\crossbarlength

\def\putdiamondbar at #1 #2 with fuzz #3 {%
  \xposition=\Xdistance{#1}
  \yposition=\Ydistance{#2}
  \dyposition=\Ydistance{#3}

\setdimensionmode
\put {\Large $\diamond$} at {\xposition} {\yposition}

\dimen0 = \yposition
  \advance \dimen0 by -\dyposition
\dimen2 = \yposition
  \advance \dimen2  by \dyposition
\putrule from {\xposition} {\dimen0}
  to {\xposition} {\dimen2}

\dimen4 = \xposition
  \advance \dimen4 by -.5\crossbarlength
\dimen6 = \xposition
  \advance \dimen6 by  .5\crossbarlength
\putrule from {\dimen4} {\dimen0} to {\dimen6} {\dimen0}
\putrule from {\dimen4} {\dimen2} to {\dimen6} {\dimen2}
\setcoordinatemode}

\newdimen\xposition
\newdimen\yposition
\newdimen\dyposition
\newdimen\crossbarlength

\def\putbigtriangledownbar at #1 #2 with fuzz #3 {%
  \xposition=\Xdistance{#1}
  \yposition=\Ydistance{#2}
  \dyposition=\Ydistance{#3}

\setdimensionmode
\put {$\bigtriangledown$} at {\xposition} {\yposition}

\dimen0 = \yposition
  \advance \dimen0 by -\dyposition
\dimen2 = \yposition
  \advance \dimen2  by \dyposition
\putrule from {\xposition} {\dimen0}
  to {\xposition} {\dimen2}

\dimen4 = \xposition
  \advance \dimen4 by -.5\crossbarlength
\dimen6 = \xposition
  \advance \dimen6 by  .5\crossbarlength
\putrule from {\dimen4} {\dimen0} to {\dimen6} {\dimen0}
\putrule from {\dimen4} {\dimen2} to {\dimen6} {\dimen2}
\setcoordinatemode}
                                           
\newdimen\xposition
\newdimen\yposition
\newdimen\dyposition
\newdimen\crossbarlength

\def\puttrianglebar at #1 #2 with fuzz #3 {%
  \xposition=\Xdistance{#1}
  \yposition=\Ydistance{#2}
  \dyposition=\Ydistance{#3}

\setdimensionmode
\put {$\triangle$} at {\xposition} {\yposition}

\dimen0 = \yposition
  \advance \dimen0 by -\dyposition
\dimen2 = \yposition
  \advance \dimen2  by \dyposition
\putrule from {\xposition} {\dimen0}
  to {\xposition} {\dimen2}

\dimen4 = \xposition
  \advance \dimen4 by -.5\crossbarlength
\dimen6 = \xposition
  \advance \dimen6 by  .5\crossbarlength
\putrule from {\dimen4} {\dimen0} to {\dimen6} {\dimen0}
\putrule from {\dimen4} {\dimen2} to {\dimen6} {\dimen2}
\setcoordinatemode}
                                          
\newdimen\xposition
\newdimen\yposition
\newdimen\dyposition
\newdimen\crossbarlength

\def\puttrianglerightbar at #1 #2 with fuzz #3 {%
  \xposition=\Xdistance{#1}
  \yposition=\Ydistance{#2}
  \dyposition=\Ydistance{#3}

\setdimensionmode
\put {\Large $\triangleright$} at {\xposition} {\yposition}

\dimen0 = \yposition
  \advance \dimen0 by -\dyposition
\dimen2 = \yposition
  \advance \dimen2  by \dyposition
\putrule from {\xposition} {\dimen0}
  to {\xposition} {\dimen2}

\dimen4 = \xposition
  \advance \dimen4 by -.5\crossbarlength
\dimen6 = \xposition
  \advance \dimen6 by  .5\crossbarlength
\putrule from {\dimen4} {\dimen0} to {\dimen6} {\dimen0}
\putrule from {\dimen4} {\dimen2} to {\dimen6} {\dimen2}
\setcoordinatemode}
                        
\newdimen\xposition
\newdimen\yposition
\newdimen\dyposition

\def\puttriangleleftbar at #1 #2 with fuzz #3 {%
  \xposition=\Xdistance{#1}
  \yposition=\Ydistance{#2}
  \dyposition=\Ydistance{#3}

\setdimensionmode
\put {\Large $\triangleleft$} at {\xposition} {\yposition}

\dimen0 = \yposition
  \advance \dimen0 by -\dyposition
\dimen2 = \yposition
  \advance \dimen2  by \dyposition
\putrule from {\xposition} {\dimen0}
  to {\xposition} {\dimen2}

\dimen4 = \xposition
  \advance \dimen4 by -.5\crossbarlength
\dimen6 = \xposition
  \advance \dimen6 by  .5\crossbarlength
\putrule from {\dimen4} {\dimen0} to {\dimen6} {\dimen0}
\putrule from {\dimen4} {\dimen2} to {\dimen6} {\dimen2}
\setcoordinatemode}

\def\beq{\begin{equation}}
\def\eeq{\end{equation}}
\def\bea{\begin{eqnarray}}
\def\eea{\end{eqnarray}}
\def\Tr{{\rm Tr}}
\def\half{ { 1 \over 2 } }

\begin{document}
\pagestyle{empty}
\begin{flushright}
NMCPP/97-17\\
hep-lat/9801009\\
\end{flushright}
\vspace*{0.01cm}
\begin{center}
{\bf\Large Confinement Artifacts in the $U(1)$ 
and $SU(2)$\\
Compact Lattice Gauge Theories\footnote{Research partially supported by
the U.S. Department of Energy}}\\
\vspace*{0.6cm}
{\bf Kevin Cahill\footnote{kevin@kevin.phys.unm.edu \quad
http://kevin.phys.unm.edu/$\tilde{\ }$kevin/} and Gary Herling\footnote{Member 
of the Center for Advanced Studies;
e-mail:~herling@unm.edu}}\\
\vspace{.5cm}
New Mexico Center for Particle Physics\\
University of New Mexico\\
Albuquerque, NM 87131-1156\\
\vspace*{1cm}
{\bf Abstract}
\end{center}
\begin{quote}
We identify the artifact
that causes confinement at strong coupling
in compact $ U(1) $ lattice gauge theory and 
show that most of the string tension
in compact $SU(2)$ lattice gauge theory is due to
plaquettes of negative trace.
\end{quote}
\vfill
\begin{flushleft}
\today\\
\end{flushleft}
\vfill\eject

\setcounter{equation}{0}
\pagestyle{plain}

\section{Introduction}
Virtually all lattice simulations are guided by 
Wilson's action~\cite{wilson}
in which the matrices of a compact group
play the role of the fields of the continuum theory.
In these simulations charged particles 
are confined at strong coupling
whether the gauge group is abelian~\cite{creutzab} 
or non-abelian~\cite{creutznab,Creu80}.
There have been a few lattice simulations 
in which the basic variables are fields.
Some of these non-compact simulations have
no exact gauge symmetry and have shown no sign
of confinement for either abelian 
or non-abelian theories~\cite{old,new}.
In one of them,
gauge invariance was partially restored
by the imposition of random gauge transformations,
and a weak confinement signal was observed~\cite{random}.
But it is not clear how much of 
that confinement signal was due
to decorrelations produced
by the noise of the imposed random gauge transformations
and how much was due to the  
attractive forces of the gluons.
Some very interesting simulations~\cite{palum,cagh} 
possess an exact lattice gauge symmetry
and display confinement for $SU(2)$ and $SU(3)$
theories but not for $U(1)$ above $\beta=0.5$\@.
The gauge fields of these simulations,
however, are not hermitian.
In all simulations, whether compact or non-compact, 
confinement has appeared only
when accompanied by significant lattice artifacts.
These lattice artifacts obscure
the confinement mechanism.\footnote{
One may even ask, with Gribov~\cite{gribov},
whether the lightness of the $u$ and $d$ quarks
is essential to confinement.}
\par
We report here the results of two studies
of lattice artifacts.
The first study shows that 
compact $U(1)$ gauge theory displays
confinement at strong coupling 
because its links and plaquettes are circles.
One may remove this artifact
by using a thin barrier to cut the circles
of the plaquettes. 
\par
The second study shows that on 
an $ 8^4 $ lattice most
of the confinement signal in compact $SU(2)$ gauge theory  
is due to plaquettes of negative trace
and that, when such plaquettes are avoided,
the ratio $ \sigma / \Lambda_L^2 $
of the string tension to 
the square of the lattice scale parameter 
is only about $ 1.6 \pm 0.2 $
as opposed to about $ 31 \pm 4 $
when they are included.
Since the avoidance of plaquettes of negative trace
is unlikely to affect perturbative quantities,
such as the lattice parameter $\Lambda_L$~\cite{Montvay,Fingberg}, 
the decrease in $ \sigma / \Lambda_L^2 $
may be attributed to a decrease in the string tension $ \sigma $\@.
Hence most of the confinement signal
observable on relatively small lattices
in compact lattice gauge theory is due to
plaquettes of negative trace. 
\par
Since our $SU(2)$ simulations were carried out
on a small lattice, we make no statement
about any possible difference 
in the continuum limit between
the physics of the standard Wilson action
and that of the positive-plaquette Wilson action.
In fact we imagine
that they are the same. 
Our result is that most of the 
confinement signal observable with
the standard Wilson action on small lattices
is due to plaquettes of negative trace,
which are lattice artifacts.
Because the lattice spacing shrinks
exponentially as the coupling becomes weaker,
we expect this result to be valid on
lattices as large as $ 100^4 $\@. 
Quark confinement has only been exhibited
on smaller lattices 
at moderately large coupling 
with actions replete with artifacts.
Our result therefore provides motivation for further
study of confinement 
on larger lattices 
at weaker coupling
with actions that suppress artifacts.

\section{Why Compact $ U(1) $ Confines}

Compact lattice simulations of $ U(1) $ gauge theory
display confinement at strong coupling.
In Figure 1 we plot the Creutz ratios\cite{Creu80} 
we obtained from simulations guided by the Wilson action
and compare them with the exact values of free
continuum QED indicated by the curves.
At $ \beta = 0.75 $, the Creutz ratios
$ \chi( 2, 2 ) $, $ \chi( 2, 3 ) $, $ \chi( 2, 4 ) $,
and $ \chi( 3, 3 ) $ all overlap
in a striking confinement signal. 
The Wilson action for $ U(1) $ confines at strong coupling
because the links 
\beq
U( x , \mu ) = \exp \left[ i \theta( x , \mu ) \right] 
= \exp \left[ i e a A_\mu( x ) \right]
\label {link}
\eeq 
take values on the unit circle.
\par
One can avoid this artifact 
by placing a thin infinite barrier
at $ | \theta | = \pi $.
We used a Metropolis algorithm 
and rejected any plaquette whose 
phase $ \theta $ was either 
greater than $ \pi - \epsilon $
or less than $ - \pi + \epsilon $.
We ran on a $ 12^4 $ lattice
and began all runs from a cold start
in which all links were unity.
Our Creutz ratios
display no sign of confinement 
and tend to follow the curves of the exact 
Creutz ratios of free continuum QED. 
The agreement with the exact ratios
is better if ones uses an effective
inverse coupling  $ \beta_e = \lambda^2 \beta $
with $ \lambda^2 = 0.73 $.
In Figure 2 we plot the measured and exact Creutz ratios 
from $ \beta = 0.25 $ to $ \beta = 1.5$\@. 
Although the continuum theory is free
and exactly soluble, 
the lattice action is non-linear and does
require renormalization.
The renormalization $ g_e = g_0 / \lambda $
is the simplest rule that in the continuum limit
satisfies 
\beq
\lim_{ g_0 \rightarrow 0 } \frac { | g^2 - g_0^2 | }{ g_0 } = 0 .
\label {ge}
\eeq
\par
We saw no confinement signal as long as the step size
was smaller than the 
thickness $ 2 \epsilon $ 
of the cut in the phase $ \theta $.
We took $ .024 < \epsilon < 0.100 $
and noticed no sensitivity to $ \epsilon $ within that range.  
The nearly overlapping points at $ \beta = 1.5 $
correspond to $ \epsilon = .024 $ and to $ \epsilon = 0.1 $\@.
We also performed simulations with Manton's action
using $ \epsilon = 0.026 $ 
and found a similar absence of confinement.
\par
Confinement signals arise when links are decorrelated.
At strong coupling many $ U(1) $ plaquettes
are near $ -1 $.  Thus when a link on a plaquette
that is near $ -1 $ is being updated, that link can
jump to a value (often far from $ 1 $)
that pushes the plaquette past the point
$ -1 $ of maximum action to a lower action.
After many such events, the links are decorrelated,
and a confinement signal appears.
\par
One may also interpret these results in terms
of monopoles~\cite{dGT}.  When the phase $\theta$ of each
plaquette is required to lie between $ \pi - \epsilon $
and $ - \pi + \epsilon $,
it follows that for $ \epsilon > 0 $,
no string can ever penetrate any plaquette.
Although the circle artifact is frequently
so interpreted,
our purpose here is rather to show that
by removing it,
we may improve the agreement between
continuum and lattice QED.

\section{Compact $ SU(2) $}

\par
In view of these results for $U(1)$,
one might wonder whether similar lattice
artifacts exist in the case of the group $SU(2)$.
Because $U(1)$ and $SU(2)$ have different
first homotopy groups ($ \pi_1( U(1) ) = Z $ but
$ \pi_1( SU(2) ) = 0 $),
it would seem likely that
the excision of a small cap 
around the antipode ($ g = -1 $) on the $ SU(2) $ group manifold 
(the three sphere $ S_3 $ in four dimensions)
would have little effect upon the Creutz ratios.
\par
To check this assumption,
we ran from cold starts
on an $ 8^4 $ lattice
and used a Metropolis algorithm
with a modified form of the Wilson action
in which we rejected plaquettes
that lay within a small cap around the antipode.
Specifically the allowed plaquettes $ P $
had to satisfy the rule
\beq
\half \Tr P > - \cos( { \pi \over 10 } ) \approx -0.951 .
\label {rule}
\eeq
In terms of the parameterization
\beq
\exp \left( i \frac { \vec \theta }{ 2 } \cdot \vec \sigma \right) 
= \cos \left( \frac { | \vec \theta | }{ 2 } \right) I
+ i \hat \theta \cdot \vec \sigma \, 
\sin \left( \frac { | \vec \theta | }{ 2 } \right) ,
\label {su2}
\eeq
the excluded cap is described by the condition
\beq
\frac{ 9 \pi }{ 5 } \le | \vec \theta |
\le \frac{ 11 \pi }{ 5 } .
\label {cond}
\eeq
The step size was small compared to 
the size of the excluded cap.
As expected,
the Creutz ratios $ \chi(i,j) $
do not depend upon whether the small cap
was excluded.
At $ \beta = 2 $, for example, the Creutz ratios
with the cap included
are: $ \chi ( 2, 2 ) = 0.5995(1) $,
$ \chi ( 2, 3 ) = 0.5813(4) $,
$ \chi ( 2, 4 ) = 0.582(5) $,
$ \chi ( 3, 3 ) = 0.556(5) $, and
$ \chi ( 3, 4 ) = 0.54(2) $;
while with the cap excluded they are:
$ \chi ( 2, 2 ) = 0.596(4) $,
$ \chi ( 2, 3 ) = 0.574(7) $,
$ \chi ( 2, 4 ) = 0.57(1) $,
$ \chi ( 3, 3 ) = 0.47(5) $, and
$ \chi ( 3, 4 ) = 0.5(1) $.
The exclusion of the cap makes little difference.
\par
But what about the excision of a large cap?
To study this question,
we performed simulations
using a Metropolis algorithm
and a modified form of the Wilson action
in which we rejected all plaquettes that had a negative trace, 
thus excluding half of the $ SU(2) $ sphere.
This positive-plaquette Wilson action
has been studied by other groups~\cite{Fingberg,others}.
We used an $ 8^4 $ lattice and
began with all $ SU(2) $ group elements 
equal to the identity (cold starts).
We used a very small step size;
the maximum change in any of the four real numbers
that describe each group element was 0.005.
Because of the small step size,
we allowed 2,000,000 sweeps for thermalization.
\par
In our positive-plaquette simulations,
the Creutz ratios $ \chi( i , j ) $ as functions
of $ \beta $ tend to follow the values given
by the $ SU(2) $ tree-level perturbative formula~\cite{Cahill}
\beq
\chi_0(i,j,\beta) = { 3 \over 2 \pi^2 \beta }
\left[ - u(i,j) - u(i-1,j-1)
+ u(i,j-1) + u(i-1,j) \right] 
\label {tree}
\eeq
where 
\begin{equation}
u(i,j) = {i \over j} \arctan{ i \over j }
+ { j \over i } \arctan{ j \over i }
- \log\left( { 1 \over i^2 } + { 1 \over j^2 } \right) .
\label {u}
\end{equation}
\par
The fit of the perturbative formula
to the data improves if we use 
an effective inverse coupling $ \beta_e = \lambda^2 \beta $
which corresponds to a renormalized coupling constant 
$ g = g_0 / \lambda $ with $ 0 < \lambda \le 1 $.
This scheme offers 
the simplest renormalization of the coupling constant
that in the continuum limit  satisfies the condition (\ref {ge}).
In Figure 3 we plot the Creutz ratios $ \chi( i , j ) $
we obtained in simulations 
guided by the positive-plaquette Wilson action. 
The curves display the perturbative values as given
by the tree-level formulas (\ref {tree}) and (\ref {u})
with the effective inverse coupling $ \beta_e = 0.54 \, \beta $.
The data follow the perturbative curves near $ \beta = 3 $
but not near $ \beta = 1 $ where the bunching of the Creutz ratios 
for different $ i $ and $ j $ indicates confinement.

\par
In the confining phase, 
the Creutz ratios $ \chi( i , j ) $
ought to be given approximately by the product
of the string tension $ \sigma $ and the 
square of the lattice spacing $ a^2( \beta_e ) $.
For the group $ SU(2) $,
the two-loop result for the dependence
of this product $ \sigma a^2( \beta_e ) $
upon the effective inverse coupling $ \beta_e $ is
\beq
\sigma a^2( \beta_e ) \approx
\frac { \sigma }{ \Lambda_L^2 } 
\exp \left[ - \frac { 6 \pi^2 \beta_e }{ 11 } 
+ \frac { 102 }{ 121 } 
\log \left( \frac { 6 \pi^2 \beta_e }{ 11 } \right) \right] ,
\label {scaling}
\eeq
where $ \Lambda_L $ is the lattice scale parameter~\cite{Creu80}\@.
We expect this scaling formula to hold
in a transition region where perturbation theory
is still valid and where the quark-antiquark
static potential $ V(r) $ is a linear combination
of a confining potential and a Coulomb potential:
\begin{equation}
\chi(i,j) = {1 \over ( \beta_{\mbox{\scriptsize max}} 
- \beta_{\mbox{\scriptsize min}} ) }
\left[ ( \beta_{\mbox{\scriptsize max}} - \beta ) \sigma a^2(\beta_e)
+ ( \beta - \beta_{\mbox{\scriptsize min}} ) \chi_0(i,j,\beta_e) \right]
\label {interpol}
\end{equation}
in which the string tension $ \sigma a^2( \beta_e ) $ is
given by the scaling formula (\ref{scaling})
and the perturbative Creutz ratio $ \chi_0(i,j, \beta_e ) $
is given by the tree-level formula (\ref{tree}--\ref{u}).
Near one end of this region, the potential
$ V(r) $ is mostly Coulomb; 
near the other end, $ V(r) $ is mostly linear.
\par
In simulations guided by the positive-plaquette Wilson action,
we found a wide transition region
between $ \beta = 1.0 $ and $ \beta = 3.0$\@. 
Our best fit to the Creutz ratios $ \chi( 2 , 2 , \beta ) $,
$ \chi( 2 , 3 , \beta ) $, $ \chi( 2 , 4 , \beta ) $,
and $ \chi( 3 , 3 , \beta ) $ at $ \beta = $ 1.0, 1.25,
1.5, 1.75, 2.0, 2.25, 2.5, 2.75, and 3.0
was $ \beta_e = 0.36 \, \beta $,
$ \beta_{\mbox{\scriptsize min}} = - 2.77 $,
$ \beta_{\mbox{\scriptsize max}} = 6.51 $, and
$ \sigma_{PP} / \Lambda_L^2 = 1.61  \pm 0.2  $. 
In the interpolation (\ref {interpol}),
the coefficient $ ( \beta_{\mbox{\scriptsize max}} - \beta ) 
/ ( \beta_{\mbox{\scriptsize max}} - \beta_{\mbox{\scriptsize min}} ) $
of the positive-plaquette 
string-tension term $ \sigma_{PP} a^2(\beta_e) $
runs from 0.59 at $ \beta = 1.0 $
to 0.38 at $ \beta = 3.0 $, and 
the coefficient $ ( \beta - \beta_{\mbox{\scriptsize min}} )
/ ( \beta_{\mbox{\scriptsize max}} - \beta_{\mbox{\scriptsize min}} ) $
of the tree-level perturbative term $ \chi_0(i,j,\beta_e) $
runs from 0.41 at $ \beta = 1.0 $ 
to 0.62 at $ \beta = 3.0 $.
Our fit to the data is displayed in Figure 4\@.
\par
The ratio $ \sigma / \Lambda_L^2 $
of the string tension to the square 
of the lattice scale parameter
is hard to measure accurately~\cite{Gutbrod}\@,
and our procedure is subject to systematic errors 
due to our measurement scheme and to the small size of our lattice.
To minimize these systematic errors, we applied the same procedure
to the unmodified Wilson action and computed
a ratio of ratios: the ratio $ \sigma_{PP} / \Lambda_L^2 $
divided by the ratio $ \sigma_W / \Lambda_L^2 $
in which $ \sigma_{PP} $ is
the positive-plaquette string tension,
$\sigma_W $ is the Wilson string tension,
and $ \Lambda_L $ is the lattice scale parameter.
\par
We performed simulations of $ SU(2) $
gauge theory guided by the unmodified Wilson
action on an $ 8^4 $ lattice and measured
the ratio $ \sigma_W / \Lambda_L^2 $ by
the same procedure that we used 
to measure the positive-plaquette Wilson action.
We found a narrow transition region
between $\beta=2.125$ and $\beta=2.5$
in which our best fit was $ \beta_e = 0.43 \, \beta $,
$ \beta_{\mbox{\scriptsize min}} = 1.42 $,
$ \beta_{\mbox{\scriptsize max}} = 2.64  $, and
$ \sigma_{W} / \Lambda_L^2 = 30.8  \pm 4  $.
In the interpolation (\ref {interpol}),
the coefficient $ ( \beta_{\mbox{\scriptsize max}} - \beta )
/ ( \beta_{\mbox{\scriptsize max}} - \beta_{\mbox{\scriptsize min}} ) $
of the Wilson string-tension term $ \sigma_{W} a^2(\beta_e) $
runs from 0.42 at $ \beta = 2.125 $
to 0.12 at $ \beta = 2.5 $, and
the coefficient $ ( \beta - \beta_{\mbox{\scriptsize min}} )
/ ( \beta_{\mbox{\scriptsize max}} - \beta_{\mbox{\scriptsize min}} ) $
of the tree-level perturbative term $ \chi_0(i,j,\beta_e) $
runs from 0.58 at $ \beta = 2.125 $
to 0.88 at $ \beta = 2.5 $.
Our fit is shown in Figure 5\@.
\par
The rejection of plaquettes that
are of negative trace 
is an entirely non-perturbative
modification of the Wilson action.
To check this assertion,
we ran a positive-plaquette simulation 
at $ \beta = 16 $
and found that the update algorithm
never tried to form a plaquette of
negative trace.
Thus for sufficiently large values of $\beta$, 
the positive-plaquette Wilson action 
is practically equivalent to the usual Wilson action.
The lattice scale parameter $ \Lambda_L $ 
is defined~\cite{Montvay} as the limit 
\beq
\Lambda_L \equiv \lim_{g\to0} 
\frac{1}{a} \exp\left( - \frac{1}{2\beta_0g^2} \right)
(\beta_0 g^2)^{-\beta_1/(2\beta_0^2)}.
\label {defofLambda}
\eeq
The value of $ \Lambda_L $
appropriate for the positive-plaquette Wilson action
is therefore the same as that for 
the unmodified Wilson action~\cite{Fingberg}.
And so our ratio of the positive-plaquette ratio
$ \sigma_{PP} / \Lambda_L^2 = 1.61 \pm 0.2  $
to the Wilsonian ratio
$ \sigma_W / \Lambda_L^2 = 30.8 \pm 4 $
may be identified as the ratio of the
positive-plaquette string tension
$ \sigma_{PP} $
to the Wilson string tension $ \sigma_{W} $,
\beq
\frac{ \sigma_{PP} / \Lambda_L^2 } { \sigma_W / \Lambda_L^2 }
= \frac{ \sigma_{PP} } { \sigma_W } 
= 0.05 .
\label{ ratio }
\eeq
Hence somewhat more than 90\% of the
confinement signal measured
in Wilsonian simulations on
an $ 8^4 $ lattice is due
to plaquettes of negative trace,
which are lattice artifacts.
\par
Because our simulations were carried out
on a small lattice,
we make no statement about any possible differences
that might exist in the continuum limit
between the physics
of the standard Wilson action and that of
the positive-plaquette Wilson action.
In fact we imagine that any such differences
are negligible,
although the approach to the continuum limit
may be smoother for the positive-plaquette action~\cite{Fingberg,others}.
Indeed beyond $ \beta = 16 $,
the Wilson action and the positive-plaquette Wilson action
are essentially equivalent.
But confinement has 
been clearly exhibited only on lattices
of moderate size ($ < 100^4 $)
at relatively large couplings
with actions that are either compact~\cite{creutznab,Creu80,booth}
or have auxiliary fields~\cite{palum,cagh}. 
Thus our result provides motivation
for further studies of confinement
at weaker coupling on larger lattices
with actions that suppress artifacts,
such as the positive-plaquette Manton action~\cite{manton}.
Such studies might suggest effective actions
suitable for use at stronger coupling. 

\section*{Acknowledgments}
\par
We have benefited particularly
from the perceptive advice of M.~Creutz.
We also are indebted to
G.~Marsaglia, W.~Press, and J.~Smit
for useful conversations,
to the Department of Energy for partial support
under grant DE-FG03-92ER40732/B005, and to
B.~Dieterle and J.~Sobolewski
for the use of computers they control.
Some of these computations were performed at 
the Albuquerque High-Performance Computing Center
and at NERSC.
Research sponsored in part by the Phillips Laboratory, 
Air Force Materiel Command, USAF, 
under cooperative agreement number F29601-93-2-0001.
The views and conclusions contained in this document are those 
of the authors and should not be interpreted as necessarily representing 
the official policies or endorsements, either expressed or implied, 
of Phillips Laboratory or the U.S. Government.


\begin{figure} [htb]
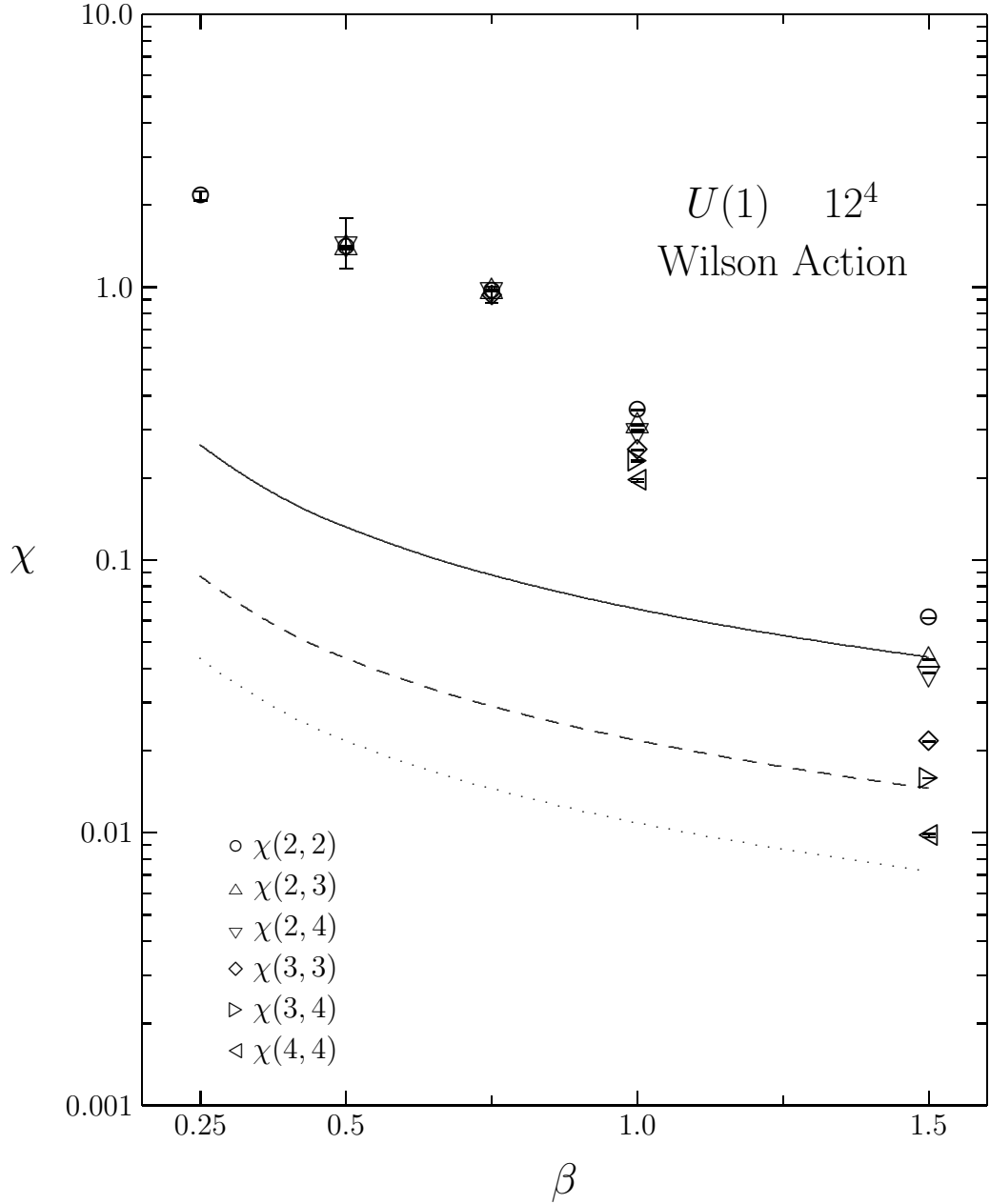

\beginpicture
\crossbarlength=5pt

\savelinesandcurves on "fig1"

\ticksin
\inboundscheckon
\setcoordinatesystem units <3.20in,1.50in>   
\setplotarea x from 0.15 to 1.6, y from -3.0 to 1.0
\axis bottom label {\Large $\beta$} ticks
  numbered at 0.25 0.5 1.0 1.5 / 
  unlabeled short from 0.75 to 0.75 by 0.25 
  from 1.25 to 1.25 by 0.25 /
\axis left  label {\Large $\chi$} 
  ticks logged 
  numbered at 0.001 0.01 0.1 1.0 10.0 /
  unlabeled short from 2.0 to 9.0 by 1.0
  from 0.2 to 0.9 by 0.1
  from 0.02 to 0.09 by 0.01 
  from 0.002 to 0.009 by 0.001 /
\axis top label {}
  ticks
  unlabeled at 0.25 0.5 1.0 1.5 / 
  unlabeled short from 0.75 to 0.75 by 0.25 
  from 1.25 to 1.25 by 0.25 /
\axis right ticks logged
  unlabeled at 0.001 0.01 0.1 1.0 10.0 /
  unlabeled short from 2.0 to 9.0 by 1.0
  from 0.2 to 0.9 by 0.1
  from 0.02 to 0.09 by 0.01 
  from 0.002 to 0.009 by 0.001 /

\put {\Large $U(1)$ \quad $12^4$} at 1.25 0.3
\put {\Large Wilson Action} at 1.25 0.1
\put {$\circ$  $\chi(2,2)$ } at 0.4 -2.05
\put {{\tiny $\triangle$}  $\chi(2,3)$ } at 0.4 -2.20
\put {{\tiny $\bigtriangledown$}  $\chi(2,4)$ } at 0.4 -2.35
\put {$\diamond$  $\chi(3,3)$ } at 0.4 -2.50
\put {$\triangleright$  $\chi(3,4)$ } at 0.4 -2.65
\put {$\triangleleft$  $\chi(4,4)$ } at 0.4 -2.80

\setquadratic
\plot 0.25 -0.57787 0.375 -0.75396 0.5 -0.87890 0.625 -0.97581
0.75 -1.05499 0.875 -1.12194 1.0 -1.1799 1.125 -1.23108
1.25 -1.27684 1.375 -1.31823 1.5 -1.3561 /  
\setdashes
\plot 0.25 -1.05909 0.375 -1.23518 0.5 -1.36012 0.625 -1.45703
0.75 -1.53621 0.875 -1.60316 1.0 -1.6611 1.125 -1.71230
1.25 -1.75806 1.375 -1.79945 1.5 -1.8371 /
\setdots
\setquadratic
\plot 0.25 -1.36123 0.375 -1.53732 0.5 -1.66226 0.625 -1.75917
0.75 -1.83835 0.875 -1.90530 1.0 -1.9634 1.125 -2.01444
1.25 -2.06020 1.375 -2.10159 1.5 -2.1397 /  
\setsolid


\inboundscheckon

\putcircbar at  0.250  0.335015 with fuzz  0.017047

\putcircbar at  0.500  0.146495 with fuzz  0.000569
\puttrianglebar at  0.500  0.144431 with fuzz  0.005735
\putbigtriangledownbar at  0.500  0.160406 with fuzz  0.091337

\putcircbar at  0.750 -0.014505 with fuzz  0.000188
\puttrianglebar at  0.750 -0.015959 with fuzz  0.001083
\putdiamondbar at  0.750 -0.032969 with fuzz  0.024641
\putbigtriangledownbar at  0.750 -0.007550 with fuzz  0.008758

\putcircbar at  1.000 -0.451865 with fuzz  0.001269
\puttrianglebar at  1.000 -0.506560 with fuzz  0.000221
\putdiamondbar at  1.000 -0.596578 with fuzz  0.000636
\putbigtriangledownbar at  1.000 -0.525971 with fuzz  0.002148
\puttrianglerightbar at  1.000 -0.636797 with fuzz  0.002288
\puttriangleleftbar at  1.000 -0.708921 with fuzz  0.005047

\putcircbar at  1.500 -1.212712 with fuzz  0.000410
\puttrianglebar at  1.500 -1.366515 with fuzz  0.000372
\putdiamondbar at  1.500 -1.666555 with fuzz  0.000863
\putbigtriangledownbar at  1.500 -1.414298 with fuzz  0.000115
\puttrianglerightbar at  1.500 -1.800294 with fuzz  0.001226
\puttriangleleftbar at  1.500 -2.011078 with fuzz  0.005996




\endpicture
\caption{The $U(1)\/$ Creutz ratios $\chi(i,j)$
as given by Wilson's action
and the exact Creutz ratios,
$\chi(2,2)$ (solid), $\chi(3,3)$ (dashes), and $\chi(4,4)$ (dots).
The confinement signal at $ \beta = 0.75 $ is striking.}
\end{figure}


\begin{figure} [htb]
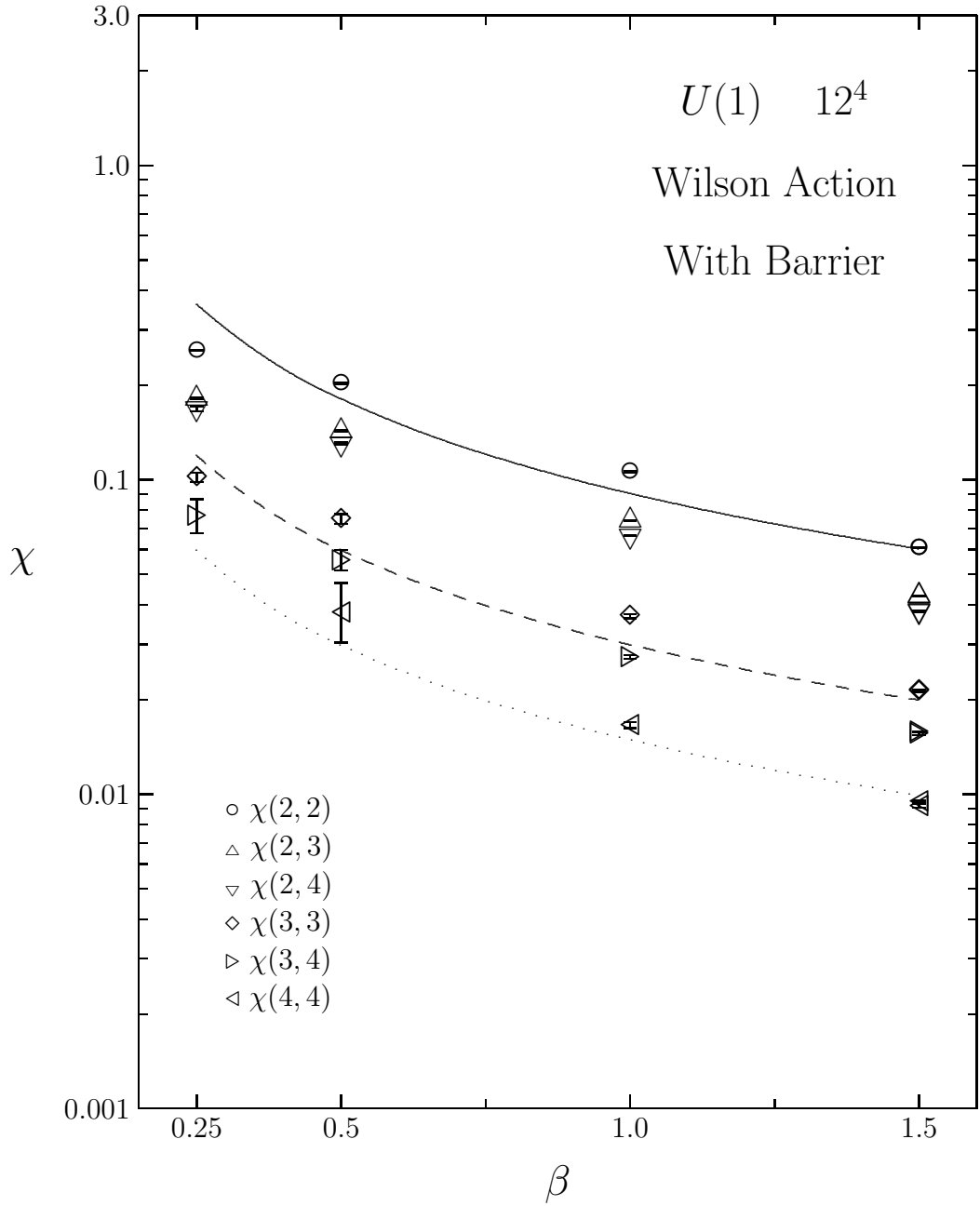

\beginpicture
\crossbarlength=5pt

\savelinesandcurves on "fig2"

\ticksin
\inboundscheckon
\setcoordinatesystem units <3.25in,1.77in>   
\setplotarea x from 0.15 to 1.6, y from -3.0 to 0.47712
\axis bottom label {\Large $\beta$} ticks
  numbered at 0.25 0.5 1.0 1.5 / 
  unlabeled short from 0.75 to 1.25 by 0.25 /
\axis left  label {\Large $\chi$} 
  ticks logged 
  numbered at 0.001 0.01 0.1 1.0 3.0 /
  unlabeled short from 2.0 to 2.0 by 1.0
  from 0.2 to 0.9 by 0.1
  from 0.02 to 0.09 by 0.01 
  from 0.002 to 0.009 by 0.001 /
\axis top label {}
  ticks
  unlabeled at 0.25 0.5 1.0 1.5 /
  unlabeled short from 0.75 to 1.25 by 0.25 /
\axis right ticks logged
  unlabeled at 0.001 0.01 0.1 1.0 /
  unlabeled short from 2.0 to 2.0 by 1.0
  from 0.2 to 0.9 by 0.1
  from 0.02 to 0.09 by 0.01
  from 0.002 to 0.009 by 0.001 /

\put {\Large $U(1)$ \quad $12^4$} at 1.25 0.2
\put {\Large Wilson Action} at 1.25 -0.05
\put {\Large With Barrier} at 1.25 -0.3
\put {$\circ$  $\chi(2,2)$ } at 0.4 -2.05
\put {{\tiny $\triangle$}  $\chi(2,3)$ } at 0.4 -2.17
\put {{\tiny $\bigtriangledown$}  $\chi(2,4)$ } at 0.4 -2.29
\put {$\diamond$  $\chi(3,3)$ } at 0.4 -2.41
\put {$\triangleright$  $\chi(3,4)$ } at 0.4 -2.53
\put {$\triangleleft$  $\chi(4,4)$ } at 0.4 -2.65

\setquadratic
\plot 0.25 -0.44119 0.375 -0.61729 0.5 -0.74222 0.625 -0.83913
0.75 -0.91832 0.875 -0.98526 1.0 -1.04325 1.125 -1.09441
1.25 -1.14016 1.375 -1.18156 1.5 -1.21935 /  
\setdashes
\plot 0.25 -0.92241 0.375 -1.09850 0.5 -1.22344 0.625 -1.32035
0.75 -1.39953 0.875 -1.46648 1.0 -1.52447 1.125 -1.57562
1.25 -1.62138 1.375 -1.66277 1.5 -1.70056 /
\setdots
\setquadratic
\plot 0.25 -1.22455 0.375 -1.40065 0.5 -1.52558 0.625 -1.62249
0.75 -1.70168 0.875 -1.76862 1.0 -1.82661 1.125 -1.87777
1.25 -1.92352 1.375 -1.96492 1.5 -2.00270 /  
\setsolid


\inboundscheckon

\putcircbar at  0.250 -0.588174 with fuzz  0.001371
\puttrianglebar at  0.250 -0.740399 with fuzz  0.002774
\putdiamondbar at  0.250 -0.992293 with fuzz  0.014903
\putbigtriangledownbar at  0.250 -0.773327 with fuzz  0.007427
\puttrianglerightbar at  0.250 -1.115541 with fuzz  0.054152

\putcircbar at  0.500 -0.692372 with fuzz  0.002113
\puttrianglebar at  0.500 -0.844554 with fuzz  0.002912
\putdiamondbar at  0.500 -1.124482 with fuzz  0.015405
\putbigtriangledownbar at  0.500 -0.885457 with fuzz  0.004124
\puttrianglerightbar at  0.500 -1.255974 with fuzz  0.032466
\puttriangleleftbar at  0.500 -1.422891 with fuzz  0.094242

\putcircbar at  1.000 -0.974224 with fuzz  0.001070
\puttrianglebar at  1.000 -1.128656 with fuzz  0.001230
\putdiamondbar at  1.000 -1.433798 with fuzz  0.006269
\putbigtriangledownbar at  1.000 -1.178238 with fuzz  0.001421
\puttrianglerightbar at  1.000 -1.563249 with fuzz  0.006167
\puttriangleleftbar at  1.000 -1.780062 with fuzz  0.009348

\putcircbar at  1.500 -1.215169 with fuzz  0.000803
\puttrianglebar at  1.500 -1.369364 with fuzz  0.001188
\putdiamondbar at  1.500 -1.671883 with fuzz  0.003234
\putbigtriangledownbar at  1.500 -1.417410 with fuzz  0.001676
\puttrianglerightbar at  1.500 -1.806340 with fuzz  0.005311
\puttriangleleftbar at  1.500 -2.036510 with fuzz  0.006109

\putcircbar at  1.500 -1.215504 with fuzz  0.000270
\puttrianglebar at  1.500 -1.369149 with fuzz  0.000081
\putdiamondbar at  1.500 -1.669131 with fuzz  0.000704
\putbigtriangledownbar at  1.500 -1.416629 with fuzz  0.000172
\puttrianglerightbar at  1.500 -1.801956 with fuzz  0.001383
\puttriangleleftbar at  1.500 -2.022757 with fuzz  0.002378




\endpicture
\caption{The $U(1)\/$ Creutz ratios $\chi(i,j)$,
as given by Wilson's action
with a small barrier,
and the exact Creutz ratios at $ \beta_e = 0.73 \, \beta $,
$\chi(2,2)$ (solid), $\chi(3,3)$ (dashes), and $\chi(4,4)$ (dots)\null.
There is no false confinement signal.}
\end{figure}

\begin{figure} [htb]
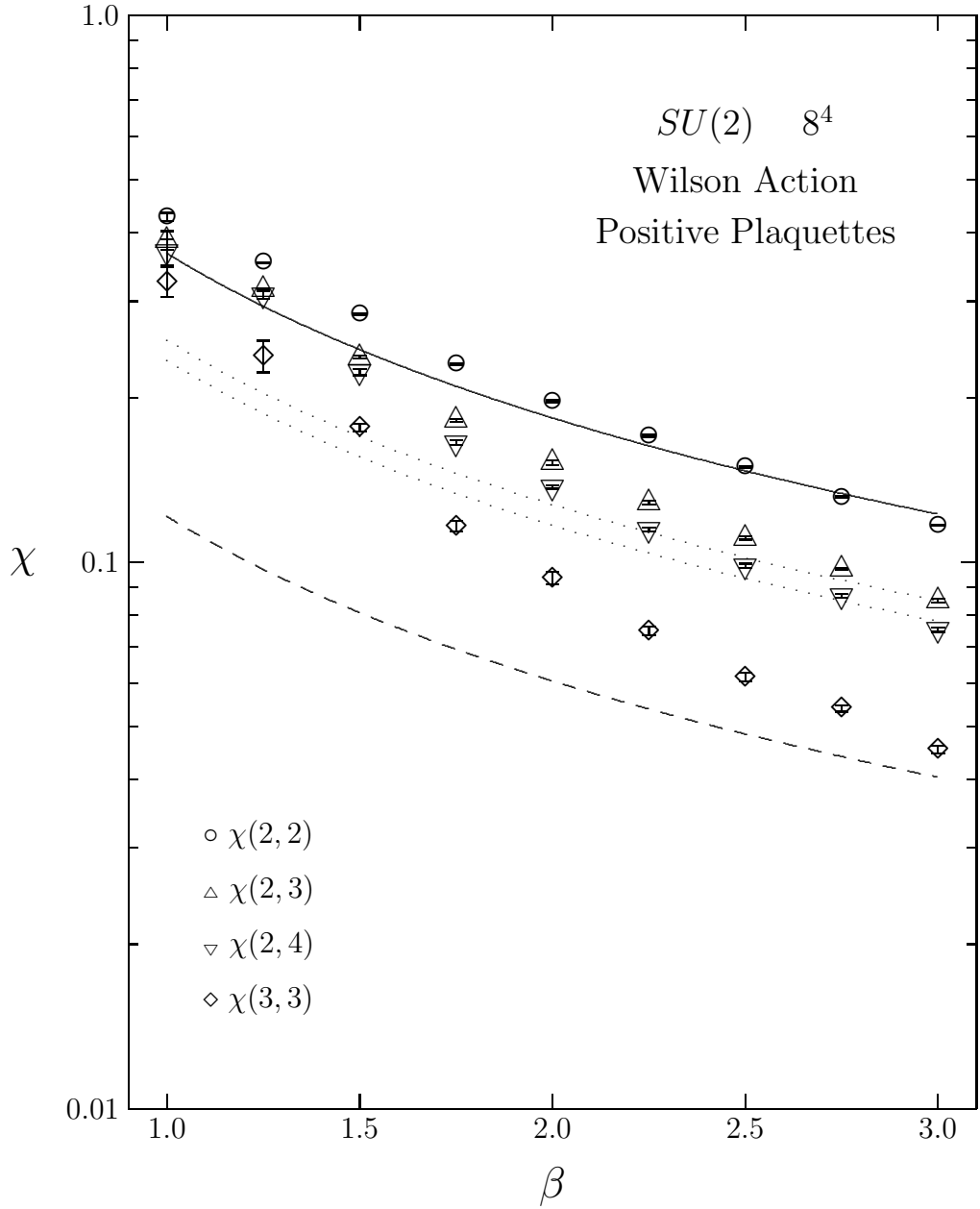

\beginpicture
\crossbarlength=5pt

\savelinesandcurves on "fig3"

\ticksin
\inboundscheckon
\setcoordinatesystem units <2.10in,2.98in>   
\setplotarea x from 0.90 to 3.1, y from -2.0 to 0.0
\axis bottom label {\Large $\beta$} ticks
  numbered at 1.0 1.5 2.0 2.5 3.0 / /
\axis left label {\Large $\chi$} 
  ticks logged 
  numbered at 0.01 0.1 1.0 /
  unlabeled short from 0.2 to 0.9 by 0.1
  from 0.02 to 0.09 by 0.01 /
\axis top label {}
  ticks
  unlabeled at 1.0 1.5 2.0 2.5 3.0 / /
\axis right ticks logged
  unlabeled at 0.01 0.1 1.0 /
  unlabeled short from 0.2 to 0.9 by 0.1
  from 0.02 to 0.09 by 0.01 /

\put {\large $SU(2)$ \quad $8^4$} at 2.5 -0.2
\put {\large Wilson Action} at 2.5 -0.3
\put {\large Positive Plaquettes} at 2.5 -0.4
\put {$\circ$  $\chi(2,2)$ } at 1.25 -1.50
\put {{\tiny $\triangle$}  $\chi(2,3)$ } at 1.25 -1.60
\put {{\tiny $\bigtriangledown$}  $\chi(2,4)$ } at 1.25 -1.70
\put {$\diamond$  $\chi(3,3)$ } at 1.25 -1.80

\inboundscheckon

\putcircbar at  1.000 -0.368756 with fuzz  0.007218
\puttrianglebar at  1.000 -0.411558 with fuzz  0.016818
\putdiamondbar at  1.000 -0.488015 with fuzz  0.027578
\putbigtriangledownbar at  1.000 -0.433632 with fuzz  0.024451

\putcircbar at  1.250 -0.452514 with fuzz  0.001273
\puttrianglebar at  1.250 -0.501653 with fuzz  0.001680
\putdiamondbar at  1.250 -0.624279 with fuzz  0.028991
\putbigtriangledownbar at  1.250 -0.511619 with fuzz  0.007016

\putcircbar at  1.500 -0.546911 with fuzz  0.001455
\puttrianglebar at  1.500 -0.624976 with fuzz  0.003079
\putdiamondbar at  1.500 -0.754526 with fuzz  0.007475
\putbigtriangledownbar at  1.500 -0.652863 with fuzz  0.005574

\putcircbar at  1.750 -0.638734 with fuzz  0.001678
\puttrianglebar at  1.750 -0.741238 with fuzz  0.002824
\putdiamondbar at  1.750 -0.934585 with fuzz  0.009355
\putbigtriangledownbar at  1.750 -0.781837 with fuzz  0.004390

\putcircbar at  2.000 -0.705873 with fuzz  0.001663
\puttrianglebar at  2.000 -0.818060 with fuzz  0.004113
\putdiamondbar at  2.000 -1.029246 with fuzz  0.011458
\putbigtriangledownbar at  2.000 -0.862459 with fuzz  0.003191

\putcircbar at  2.250 -0.769155 with fuzz  0.001909
\puttrianglebar at  2.250 -0.891108 with fuzz  0.002990
\putdiamondbar at  2.250 -1.126196 with fuzz  0.008294
\putbigtriangledownbar at  2.250 -0.940274 with fuzz  0.003309

\putcircbar at  2.500 -0.826721 with fuzz  0.001554
\puttrianglebar at  2.500 -0.956099 with fuzz  0.002947
\putdiamondbar at  2.500 -1.210610 with fuzz  0.008086
\putbigtriangledownbar at  2.500 -1.007126 with fuzz  0.004285

\putcircbar at  2.750 -0.881185 with fuzz  0.001306
\puttrianglebar at  2.750 -1.012706 with fuzz  0.001463
\putdiamondbar at  2.750 -1.267740 with fuzz  0.006074
\putbigtriangledownbar at  2.750 -1.062349 with fuzz  0.003296

\putcircbar at  3.000 -0.932790 with fuzz  0.001418
\puttrianglebar at  3.000 -1.071015 with fuzz  0.003574
\putdiamondbar at  3.000 -1.343040 with fuzz  0.007246
\putbigtriangledownbar at  3.000 -1.123609 with fuzz  0.004098



\setsolid
\setquadratic 
\plot
 1.00    -0.435204
 1.10    -0.476596
 1.20    -0.514385
 1.30    -0.549147
 1.40    -0.581332
 1.50    -0.611295
 1.60    -0.639324
 1.70    -0.665653
 1.80    -0.690476
 1.90    -0.713957
 2.00    -0.736234
 2.10    -0.757423
 2.20    -0.777626
 2.30    -0.796932
 2.40    -0.815415
 2.50    -0.833144
 2.60    -0.850177
 2.70    -0.866568
 2.80    -0.882362
 2.90    -0.897602
 3.00    -0.912325
/
\setdots
\setquadratic 
\plot
 1.00    -0.593843
 1.10    -0.635236
 1.20    -0.673024
 1.30    -0.707786
 1.40    -0.739971
 1.50    -0.769934
 1.60    -0.797963
 1.70    -0.824292
 1.80    -0.849115
 1.90    -0.872596
 2.00    -0.894873
 2.10    -0.916062
 2.20    -0.936266
 2.30    -0.955571
 2.40    -0.974054
 2.50    -0.991783
 2.60    -1.008816
 2.70    -1.025207
 2.80    -1.041001
 2.90    -1.056241
 3.00    -1.070964
/
\setdots
\setquadratic 
\plot
 1.00    -0.631339
 1.10    -0.672732
 1.20    -0.710520
 1.30    -0.745283
 1.40    -0.777467
 1.50    -0.807430
 1.60    -0.835459
 1.70    -0.861788
 1.80    -0.886612
 1.90    -0.910093
 2.00    -0.932369
 2.10    -0.953559
 2.20    -0.973762
 2.30    -0.993067
 2.40    -1.011550
 2.50    -1.029279
 2.60    -1.046313
 2.70    -1.062703
 2.80    -1.078497
 2.90    -1.093737
 3.00    -1.108460
/
\setdashes
\setquadratic 
\plot
 1.00    -0.916421
 1.10    -0.957814
 1.20    -0.995602
 1.30    -1.030364
 1.40    -1.062549
 1.50    -1.092512
 1.60    -1.120541
 1.70    -1.146870
 1.80    -1.171694
 1.90    -1.195175
 2.00    -1.217451
 2.10    -1.238640
 2.20    -1.258844
 2.30    -1.278149
 2.40    -1.296632
 2.50    -1.314361
 2.60    -1.331394
 2.70    -1.347785
 2.80    -1.363579
 2.90    -1.378819
 3.00    -1.393542
/

\endpicture
\caption{The $SU(2)\/$ Creutz ratios $\chi(i,j)$
as given by Wilson's action
restricted to positive plaquettes
and the tree-level perturbative formula $ \chi_0( i, j, \beta_e ) $
for the Creutz ratios
$\chi(2,2)$ (solid), $\chi(2,3)$ (dots), $\chi(2,4)$ (dots),
and $\chi(3,3)$ (dashes) at the effective
inverse coupling $ \beta_e = 0.54 \, \beta $.} 
\end{figure}


\begin{figure} [htb]
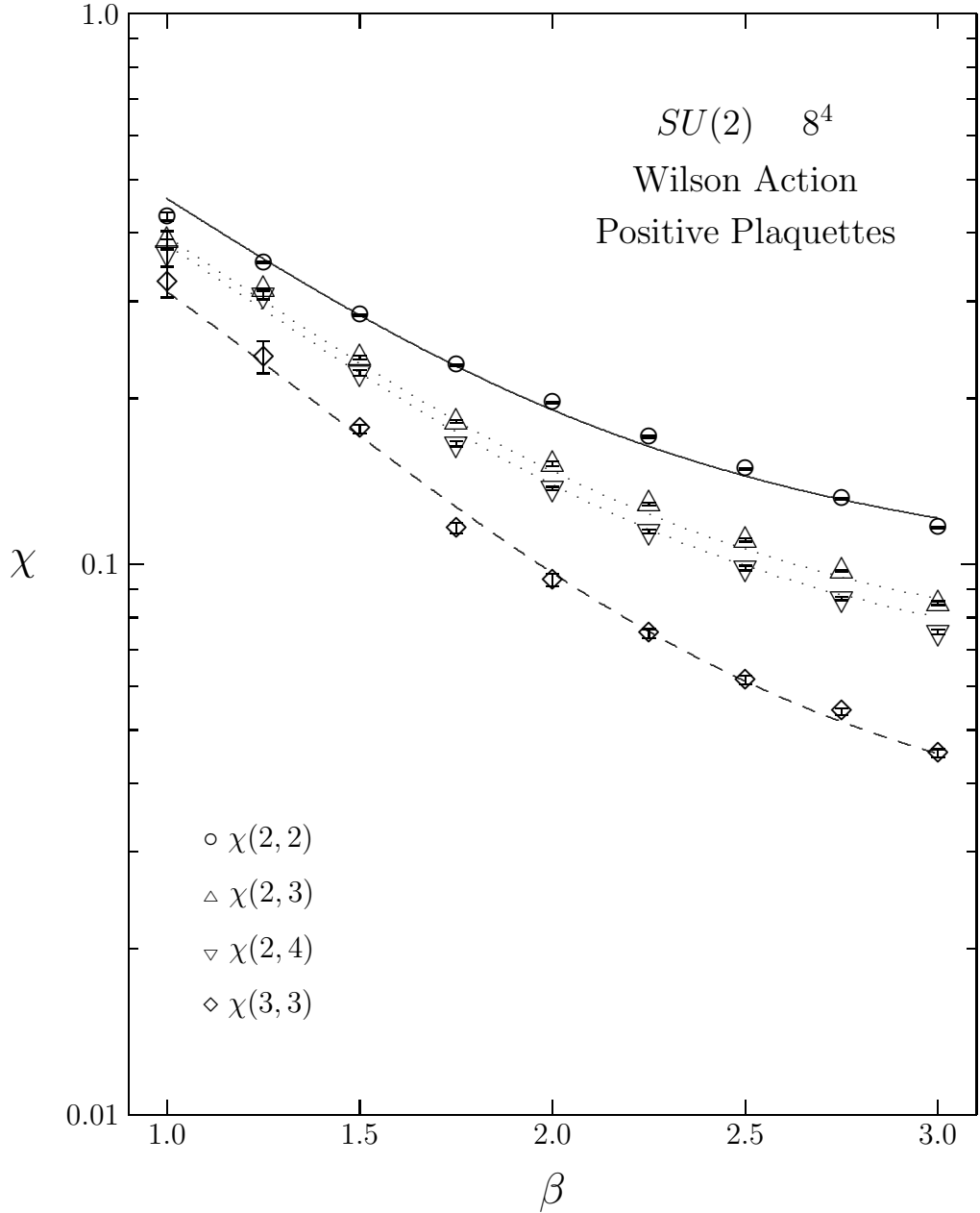

\beginpicture
\crossbarlength=5pt

\savelinesandcurves on "fig4"

\ticksin
\inboundscheckon
\setcoordinatesystem units <2.10in,3.00in>   
\setplotarea x from 0.90 to 3.1, y from -2.0 to 0.0
\axis bottom label {\Large $\beta$} ticks
  numbered at 1.0 1.5 2.0 2.5 3.0 / /
\axis left label {\Large $\chi$} 
  ticks logged 
  numbered at 0.01 0.1 1.0 /
  unlabeled short from 0.2 to 0.9 by 0.1
  from 0.02 to 0.09 by 0.01 /
\axis top label {}
  ticks
  unlabeled at 1.0 1.5 2.0 2.5 3.0 / /
\axis right ticks logged
  unlabeled at 0.01 0.1 1.0 /
  unlabeled short from 0.2 to 0.9 by 0.1
  from 0.02 to 0.09 by 0.01 /

\put {\large $SU(2)$ \quad $8^4$} at 2.5 -0.2
\put {\large Wilson Action} at 2.5 -0.3
\put {\large Positive Plaquettes} at 2.5 -0.4
\put {$\circ$  $\chi(2,2)$ } at 1.25 -1.50
\put {{\tiny $\triangle$}  $\chi(2,3)$ } at 1.25 -1.60
\put {{\tiny $\bigtriangledown$}  $\chi(2,4)$ } at 1.25 -1.70
\put {$\diamond$  $\chi(3,3)$ } at 1.25 -1.80


\inboundscheckon


\putcircbar at  1.000 -0.368756 with fuzz  0.007218
\puttrianglebar at  1.000 -0.411558 with fuzz  0.016818
\putdiamondbar at  1.000 -0.488015 with fuzz  0.027578
\putbigtriangledownbar at  1.000 -0.433632 with fuzz  0.024451

\putcircbar at  1.250 -0.452514 with fuzz  0.001273
\puttrianglebar at  1.250 -0.501653 with fuzz  0.001680
\putdiamondbar at  1.250 -0.624279 with fuzz  0.028991
\putbigtriangledownbar at  1.250 -0.511619 with fuzz  0.007016

\putcircbar at  1.500 -0.546911 with fuzz  0.001455
\puttrianglebar at  1.500 -0.624976 with fuzz  0.003079
\putdiamondbar at  1.500 -0.754526 with fuzz  0.007475
\putbigtriangledownbar at  1.500 -0.652863 with fuzz  0.005574

\putcircbar at  1.750 -0.638734 with fuzz  0.001678
\puttrianglebar at  1.750 -0.741238 with fuzz  0.002824
\putdiamondbar at  1.750 -0.934585 with fuzz  0.009355
\putbigtriangledownbar at  1.750 -0.781837 with fuzz  0.004390

\putcircbar at  2.000 -0.705873 with fuzz  0.001663
\puttrianglebar at  2.000 -0.818060 with fuzz  0.004113
\putdiamondbar at  2.000 -1.029246 with fuzz  0.011458
\putbigtriangledownbar at  2.000 -0.862459 with fuzz  0.003191

\putcircbar at  2.250 -0.769155 with fuzz  0.001909
\puttrianglebar at  2.250 -0.891108 with fuzz  0.002990
\putdiamondbar at  2.250 -1.126196 with fuzz  0.008294
\putbigtriangledownbar at  2.250 -0.940274 with fuzz  0.003309

\putcircbar at  2.500 -0.826721 with fuzz  0.001554
\puttrianglebar at  2.500 -0.956099 with fuzz  0.002947
\putdiamondbar at  2.500 -1.210610 with fuzz  0.008086
\putbigtriangledownbar at  2.500 -1.007126 with fuzz  0.004285

\putcircbar at  2.750 -0.881185 with fuzz  0.001306
\puttrianglebar at  2.750 -1.012706 with fuzz  0.001463
\putdiamondbar at  2.750 -1.267740 with fuzz  0.006074
\putbigtriangledownbar at  2.750 -1.062349 with fuzz  0.003296

\putcircbar at  3.000 -0.932790 with fuzz  0.001418
\puttrianglebar at  3.000 -1.071015 with fuzz  0.003574
\putdiamondbar at  3.000 -1.343040 with fuzz  0.007246
\putbigtriangledownbar at  3.000 -1.123609 with fuzz  0.004098


\setsolid
\setquadratic 
\plot
 1.00    -0.335613
 1.10    -0.379774
 1.20    -0.423328
 1.30    -0.465976
 1.40    -0.507435
 1.50    -0.547450
 1.60    -0.585802
 1.70    -0.622308
 1.80    -0.656828
 1.90    -0.689266
 2.00    -0.719568
 2.10    -0.747719
 2.20    -0.773742
 2.30    -0.797693
 2.40    -0.819654
 2.50    -0.839728
 2.60    -0.858033
 2.70    -0.874698
 2.80    -0.889855
 2.90    -0.903637
 3.00    -0.916173
/
\setdots
\setquadratic 
\plot
 1.00    -0.404886
 1.10    -0.451539
 1.20    -0.498190
 1.30    -0.544446
 1.40    -0.589938
 1.50    -0.634334
 1.60    -0.677330
 1.70    -0.718665
 1.80    -0.758118
 1.90    -0.795510
 2.00    -0.830709
 2.10    -0.863631
 2.20    -0.894235
 2.30    -0.922525
 2.40    -0.948543
 2.50    -0.972364
 2.60    -0.994090
 2.70    -1.013844
 2.80    -1.031763
 2.90    -1.047992
 3.00    -1.062678
/
\setdots
\setquadratic 
\plot
 1.00    -0.419209
 1.10    -0.466431
 1.20    -0.513794
 1.30    -0.560888
 1.40    -0.607328
 1.50    -0.652763
 1.60    -0.696878
 1.70    -0.739391
 1.80    -0.780064
 1.90    -0.818699
 2.00    -0.855145
 2.10    -0.889298
 2.20    -0.921101
 2.30    -0.950541
 2.40    -0.977647
 2.50    -1.002485
 2.60    -1.025149
 2.70    -1.045759
 2.80    -1.064450
 2.90    -1.081367
 3.00    -1.096663
/
\setdashes
\setquadratic 
\plot
 1.00    -0.504845
 1.10    -0.555908
 1.20    -0.608138
 1.30    -0.661036
 1.40    -0.714145
 1.50    -0.767046
 1.60    -0.819347
 1.70    -0.870682
 1.80    -0.920709
 1.90    -0.969114
 2.00    -1.015614
 2.10    -1.059961
 2.20    -1.101951
 2.30    -1.141423
 2.40    -1.178268
 2.50    -1.212427
 2.60    -1.243892
 2.70    -1.272702
 2.80    -1.298938
 2.90    -1.322716
 3.00    -1.344181
/

\endpicture
\caption{The $SU(2)\/$ Creutz ratios $\chi(i,j)$
as given by Wilson's action
restricted to positive plaquettes
and the fit (\ref {interpol}) to the Creutz ratios
$\chi(2,2)$ (solid), $\chi(2,3)$ (dots), $\chi(2,4)$
(dots), and $\chi(3,3)$ (dashes).}
\end{figure}


\begin{figure} [htb]
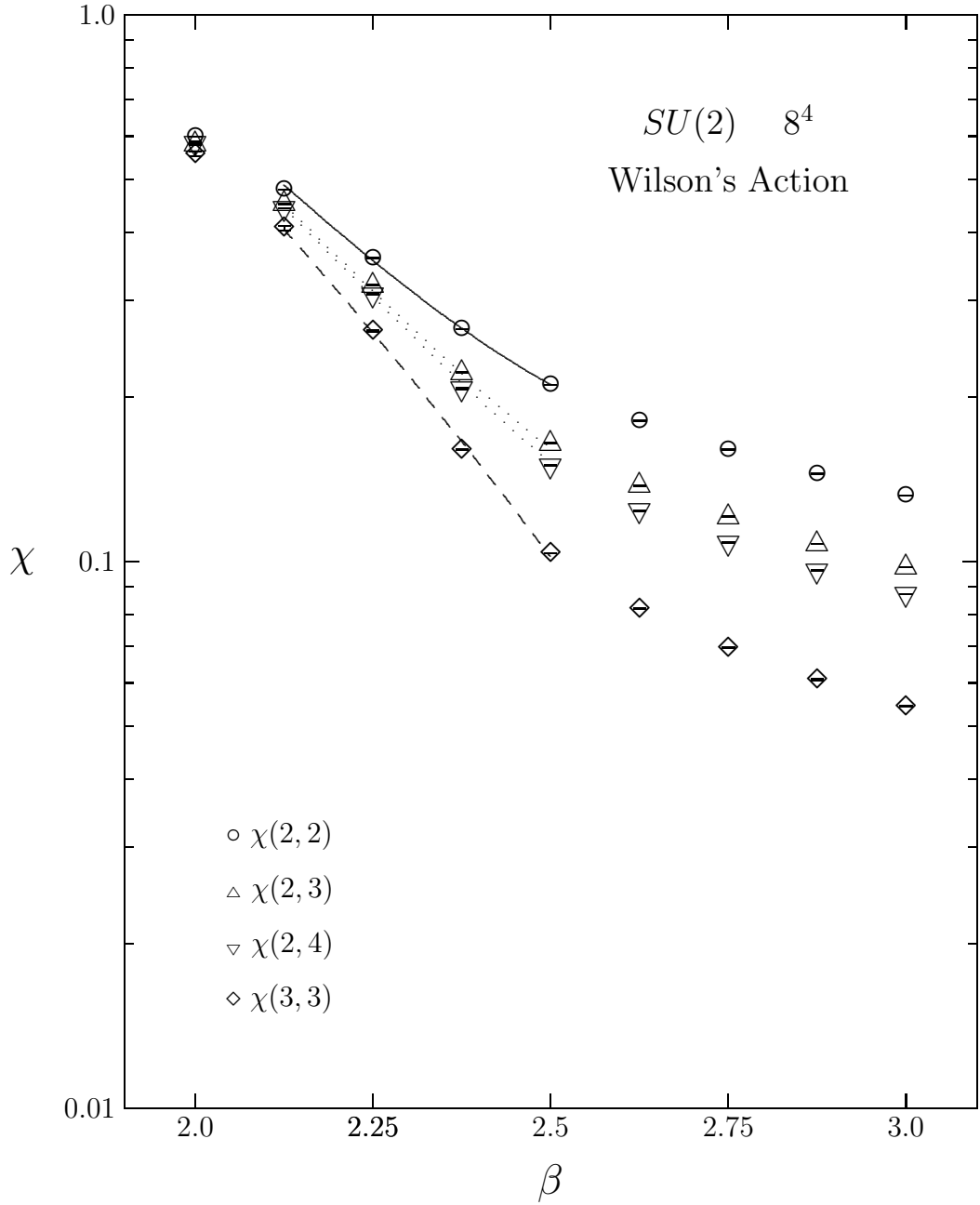

\beginpicture
\crossbarlength=5pt

\savelinesandcurves on "fig5"

\ticksin
\inboundscheckon
\setcoordinatesystem units <4.00in,3.08in>   
\setplotarea x from 1.90 to 3.1, y from -2.0 to 0.0
\axis bottom label {\Large $\beta$} ticks
  numbered at 2.0 2.25 2.5 2.25 2.75 3.0 / /
\axis left label {\Large $\chi$} 
  ticks logged 
  numbered at 0.01 0.1 1.0 /
  unlabeled short from 0.2 to 0.9 by 0.1
  from 0.02 to 0.09 by 0.01 /
\axis top label {}
  ticks
  unlabeled at 2.0 2.25 2.5 2.75 3.0 / /
\axis right ticks logged
  unlabeled at 0.01 0.1 1.0 /
  unlabeled short from 0.2 to 0.9 by 0.1
  from 0.02 to 0.09 by 0.01 /

\put {\large $SU(2)$ \quad $8^4$} at 2.75 -0.2
\put {\large Wilson's Action} at 2.75 -0.3
\put {$\circ$  $\chi(2,2)$ } at 2.125 -1.5
\put {{\tiny $\triangle$}  $\chi(2,3)$ } at 2.125 -1.60
\put {{\tiny $\bigtriangledown$}  $\chi(2,4)$ } at 2.125 -1.7
\put {$\diamond$  $\chi(3,3)$ } at 2.125 -1.80


\inboundscheckon

\putcircbar at  2.000 -0.222202 with fuzz  0.000090
\puttrianglebar at  2.000 -0.235597 with fuzz  0.000316
\putdiamondbar at  2.000 -0.254615 with fuzz  0.003947
\putbigtriangledownbar at  2.000 -0.235062 with fuzz  0.003373

\putcircbar at  2.125 -0.319638 with fuzz  0.000133
\puttrianglebar at  2.125 -0.346208 with fuzz  0.001165
\putdiamondbar at  2.125 -0.390195 with fuzz  0.004188
\putbigtriangledownbar at  2.125 -0.354089 with fuzz  0.000202

\putcircbar at  2.250 -0.444761 with fuzz  0.000191
\puttrianglebar at  2.250 -0.494239 with fuzz  0.000341
\putdiamondbar at  2.250 -0.578092 with fuzz  0.000721
\putbigtriangledownbar at  2.250 -0.511263 with fuzz  0.000599

\putcircbar at  2.375 -0.574835 with fuzz  0.000009
\puttrianglebar at  2.375 -0.654512 with fuzz  0.000312
\putdiamondbar at  2.375 -0.795109 with fuzz  0.000177
\putbigtriangledownbar at  2.375 -0.683864 with fuzz  0.001294

\putcircbar at  2.500 -0.676461 with fuzz  0.000168
\puttrianglebar at  2.500 -0.783176 with fuzz  0.000288
\putdiamondbar at  2.500 -0.984805 with fuzz  0.000673
\putbigtriangledownbar at  2.500 -0.824443 with fuzz  0.000382

\putcircbar at  2.625 -0.742496 with fuzz  0.000019
\puttrianglebar at  2.625 -0.860713 with fuzz  0.000447
\putdiamondbar at  2.625 -1.086435 with fuzz  0.000346
\putbigtriangledownbar at  2.625 -0.906951 with fuzz  0.000218

\putcircbar at  2.750 -0.794566 with fuzz  0.000338
\puttrianglebar at  2.750 -0.918308 with fuzz  0.000485
\putdiamondbar at  2.750 -1.157400 with fuzz  0.000471
\putbigtriangledownbar at  2.750 -0.965765 with fuzz  0.000666

\putcircbar at  2.875 -0.839278 with fuzz  0.000021
\puttrianglebar at  2.875 -0.967614 with fuzz  0.000101
\putdiamondbar at  2.875 -1.216402 with fuzz  0.001390
\putbigtriangledownbar at  2.875 -1.016410 with fuzz  0.000129

\putcircbar at  3.000 -0.878740 with fuzz  0.000158
\puttrianglebar at  3.000 -1.010459 with fuzz  0.000290
\putdiamondbar at  3.000 -1.264739 with fuzz  0.000053
\putbigtriangledownbar at  3.000 -1.059818 with fuzz  0.000281



\setsolid
\setquadratic 
\plot
 2.125    -0.311351
 2.144    -0.332628
 2.163    -0.353730
 2.181    -0.374635
 2.200    -0.395321
 2.219    -0.415765
 2.237    -0.435942
 2.256    -0.455827
 2.275    -0.475396
 2.294    -0.494621
 2.313    -0.513476
 2.331    -0.531934
 2.350    -0.549968
 2.369    -0.567550
 2.388    -0.584654
 2.406    -0.601254
 2.425    -0.617324
 2.444    -0.632840
 2.462    -0.647779
 2.481    -0.662120
 2.500    -0.675843
/
\setdots
\setquadratic 
\plot
 2.125    -0.346754
 2.144    -0.370573
 2.163    -0.394365
 2.181    -0.418114
 2.200    -0.441803
 2.219    -0.465413
 2.237    -0.488924
 2.256    -0.512315
 2.275    -0.535562
 2.294    -0.558640
 2.313    -0.581523
 2.331    -0.604183
 2.350    -0.626589
 2.369    -0.648711
 2.388    -0.670515
 2.406    -0.691968
 2.425    -0.713035
 2.444    -0.733678
 2.462    -0.753861
 2.481    -0.773546
 2.500    -0.792695
/
\setdots
\setquadratic 
\plot
 2.125    -0.353730
 2.144    -0.378076
 2.163    -0.402430
 2.181    -0.426778
 2.200    -0.451105
 2.219    -0.475394
 2.237    -0.499626
 2.256    -0.523782
 2.275    -0.547840
 2.294    -0.571777
 2.313    -0.595568
 2.331    -0.619185
 2.350    -0.642601
 2.369    -0.665783
 2.388    -0.688701
 2.406    -0.711319
 2.425    -0.733602
 2.444    -0.755511
 2.462    -0.777009
 2.481    -0.798055
 2.500    -0.818607
/
\setdashes
\setquadratic 
\plot
 2.125    -0.392983
 2.144    -0.420474
 2.163    -0.448211
 2.181    -0.476198
 2.200    -0.504437
 2.219    -0.532933
 2.237    -0.561687
 2.256    -0.590703
 2.275    -0.619983
 2.294    -0.649528
 2.313    -0.679339
 2.331    -0.709415
 2.350    -0.739756
 2.369    -0.770359
 2.388    -0.801220
 2.406    -0.832333
 2.425    -0.863692
 2.444    -0.895285
 2.462    -0.927100
 2.481    -0.959121
 2.500    -0.991329
/

\endpicture
\caption{The $SU(2)\/$ Creutz ratios $\chi(i,j)$
as given by the unmodified Wilson action
and the fit (\ref {interpol}) to the Creutz ratios
$\chi(2,2)$ (solid), $\chi(2,3)$ (dots), $\chi(2,4)$
(dots), and $\chi(3,3)$ (dashes) between $\beta=2.125$
and $\beta=2.5$.}
\end{figure}

\end{document}